\documentclass[jphysg,12pt]{article}
\usepackage[dvips]{graphics}
\usepackage{epsfig}
\newcommand{\be}{\begin{equation}}
\newcommand{\ee}{\end{equation}}
\def\lapproxeq{\lower .7ex\hbox{$\;\stackrel{\textstyle<}{\sim}\;$}}

\begin{document}
\baselineskip 20pt plus .1pt  minus .1pt
\pagestyle{plain}
\voffset -2.0cm
\hoffset -1.0cm
\setcounter{page}{01}
\rightline {}
\vskip 1.0cm

\begin{center}
{\Large {\bf The Origin of Cosmic Rays}}
\end{center}
\begin{center}
A.D.Erlykin $^{1,2}$, A.W.Wolfendale $^2$
\end{center}

\begin{flushleft}
\small (1) P. N. Lebedev Physical Institute, Moscow, Russia\\
(2) Department of Physics, University of Durham, Durham, UK
\end{flushleft}

\begin{abstract}
It is generally regarded that the bulk of cosmic rays originate in the Galaxy and
that those below the 'knee' (~the rapid steepening in the energy
spectrum~) at a few PeV come from Galactic supernovae, the
particles being accelerated by the shocks in the supernova remnants. At higher
energies, there are problems in that conventional SNR - which surely constitute the
bulk of the sources - have a natural limit at a few tens of PeV ( for iron nuclei ).
The question of the origin of particles above this limit is thus an open one. Here we
examine a number of possibilities: a variety of supernovae and hypernovae, pulsars,
a Giant Galactic Halo and an Extragalactic origin.

A relevant property of any model is the extent to which it can
provide the lack of significant irregularity of the energy
spectrum, its intensity and shape together with structures such as
the 'second knee' at the sub-EeV energy, in addition to the well
known 'knee' and 'ankle'. Although it is appreciated that spectral
measurements are subject to systematic as well as random errors we
consider that contemporary data are good enough to allow at least
some progress in this new field.  These aspects are examined for
particles of all energies and it is shown that they can constrain
some parameters of the proposed models.

In the search for origin above PeV energies we conclude that
shocks in the Galactic Halo, whatever their source (~Galactic
wind, relativistic plasmoids - 'cannonballs', multiple shocks from
supernovae etc.~) are most likely, pulsars such as B0656+14 and
hypernovae come a close second although such a suggestion is not
without its difficulties. What is most important is that {\em
trapping of particles in the Halo
 is needed to reduce irregularities
of the energy spectra both below and above the 'knee' caused by the stochastic nature
of supernova explosions and other potential (~discrete~) Galactic sources}.

We argue that precise experimental studies of spectral
`irregularities` will provide considerable help in the search for
cosmic ray origin.

\noindent
\end{abstract}

\section{Introduction}
Some ninety years after the discovery of the cosmic radiation (CR) there is still
dispute about its origin, at least for energies above the 'knee' in the spectrum.
Indeed, it has often been said that 'had such particles not been observed they would
never have been predicted'.

As with other high energy bands in the CR spectrum, the problem is not with the total
energy content but rather with achieving single particle energies of the required
magnitude.

Below the knee at 3 PeV there is general agreement that supernova remnants (SNR) are
responsible and much work has been done \cite{Hillas,Haung1}. We, ourselves, have
published a number of
papers ({\em eg} \cite{EW1,EW2}) and, most particularly \cite{EW3,EW4}, in which we
have made
extensive Monte Carlo calculations for SN distributed at random in the Galaxy and
determined the expected spectral shape. A satisfactory model has resulted, for
energies below about 10 PeV, a characteristic feature being the presence of fine
structure, viz. 'peaks' in the spectrum at about 3 and 12 PeV (~when the spectrum is
plotted as $E^3I(E)$~) due, in our initial concept,
to oxygen and iron nuclei (~helium and oxygen not being excluded, however~) accelerated
 by a single, recent and nearby SNR. Very recent developments \cite{Ostrow,Thors} have
lent support to this model.

In what follows, we examine the source of CR at higher energies in
some detail. In fact, for some of the analysis, the proposed
mechanisms for high energy particles impinge on the lower energies
and it is necessary to examine the lower energy region as well.
Particularly important is the relevance of the observed curvature
in the spectrum of atomic nuclei to the model of shock
acceleration and the fluctuations in spectral shape (~see the
Appendix~). Potential sources for post-PeV particles are various
supernovae and hypernovae, pulsars, Galactic Halo shocks,
acceleration in the magnified magnetic fields of young supernova
remnants (~SNR~), re-acceleration by collective effects of
multiple SNR shocks and Extragalactic
 (~EG~) phenomena.  We endeavour to rank these possibilities in
 decreasing order of likelihood.

\section{Phenomenology of Cosmic Rays above PeV energies}

At the knee at $\sim$3 PeV the cosmic ray energy spectrum suddenly
steepens and continues to fall with a slope index of about 3.0 up
to $\sim$400 PeV. It is commonly assumed that the knee is the only
feature in the energy spectrum until the `ankle' is reached, at an
energy approaching 1000PeV.  It is, however, in our view, highly likely
that there is more `structure' in the spectral shape; every
physical phenomena in the Galaxy is spatially (and temporally)
variable  i.e., is irregular and there is no reason why the CR
spectrum should not have irregularities.  In most - and perhaps
all - models of the knee there should be fine structure at some
level in the general power law, due to cut-offs in the spectra of
constituent CR nuclei. Indeed, we have claimed to have found a
small second peak in the all-particle spectrum at about 12 PeV
\cite{EW5}. If the dominant element at the knee is oxygen, then
the position of the second peak can
 be an indicator that it is due to the cut-off of the iron group. Since the abundance
of trans-iron elements is small, CR above $\sim$20 PeV should have another origin
although there is a model in which trans-iron elements can make a substantial
contribution to the total CR intensity after the cut-off of iron \cite{Hoer1}; a model
to which we do not subscribe.

There is another interpretation for the structure of the spectrum at and above the
knee: that rather than oxygen, it is, instead, helium which is dominant at the knee,
and the second peak is attributed to oxygen (~or in general to the CNO group~) rather
than iron. In this case, if the iron is
still abundant in CR, one can expect the existence of a third peak in the spectrum at
$\sim$40 PeV caused by the cut-off of iron, assuming that all cut-offs occur at the
same rigidity. The experimental indications of the
existence of a third peak are gradually accumulating, but the general situation is
still controversial \cite{Kamp,EW6}. Hopes of clarifying the situation are connected
with the expected results from KASCADE-Grande \cite{Haung2}.

The positions of the second and third peaks are practically independent of the model of
 the knee. In two alternatives, in which cut-offs of the individual elements occur at
a fixed rigidity or at a fixed energy per nucleon, the position of the cut-off
energy in the energy spectrum is proportional either to the charge $Z$ or to the mass
$A$ of the individual element. In both cases the interval between the cut-off energy
for He and Fe is practically the same (~in logarithmic units~).

Independently of the visibility of the third peak it is evident
from the experimental data that above 40 PeV ($logE, GeV = 7.6$)
we have a wide energy interval in which the CR energy spectrum
extends as a power law with an approximately constant differential
slope index of $\gamma \simeq$ 3.04 up to $logE, GeV = 8.6-8.9$.
At this energy the CR spectrum has another faint feature known as
'the second knee', which is a further steepening with $\gamma
\simeq $3.18 up to an energy of the 'ankle' at $logE \approx
9.5-10.0$. Berezinsky et al. \cite{Bere1,Bere2} consider that soon
after the third 'iron' peak extragalactic protons enter the scene
and the second knee arises from them with the rise of their energy
loss for the production of electron-positron pairs in
electromagnetic interactions with the cosmic microwave background
(~CMB~) causing the ankle. In this case the good tuning of
Galactic switch-off and Extragalactic switch-on energies in the
region of $logE = 8-9$ is puzzling.

On the contrary, Wibig and Wolfendale \cite{Wibig} argue that
Galactic CR extend further after $logE = 7.6$ up to the ankle, but
they do not consider the origin of these Galactic CR, which should
certainly change after the termination of iron from conventional 
supernovae. The hypothetical
contribution of trans-iron elements \cite{Hoer1} cannot completely
fill the gap between the iron peak and the ankle. Here we shall
try to analyze the situation at higher energies, largely in the
PeV-EeV interval between the knee and the ankle, starting with
supernovae and hypernovae, sources which have relevance both above
and below the knee.

\section{Analysis of the different models of CR origin above the knee}

\subsection{A variety of supernovae and hypernovae}

To our knowledge, it was P.L.Biermann who suggested that another
kind of supernova, coming from more massive and hot stars, such as
Wolf-Rayet stars, might be responsible for the production of CR
with energies above the knee \cite{Bier1}. This idea has been
developed by Sveshnikova \cite{Svesh}, who used the evidence that
there are different classes of SN and even within a single class
there is a spread of explosion energies. There is also a weak
dependence of the maximum energy of accelerated particles
$E_{max}$ on the explosion energy within a single class but the
difference between $E_{max}$ for different classes of SN might be
higher. The author integrated the spectra of accelerated CR with
assumed distributions of explosion energies and showed that with
certain fractions of different SN classes, taken mainly from
observations, and certain assumptions about $E_{max}$ for
different classes, it is possible to reproduce the shape of the CR
spectrum above the knee. As would be expected, the dominant
contribution to the spectrum comes from the most energetic
classes: SNIbc and SNIIn.

We have made calculations for a number of scenarios, starting with
the `standard SN', viz SN energy of $10^{51}$ ergs, of which
10$\%$ goes into CR and a differential energy spectrum of the form 
$E^{-2.15}$.  Full details of the diffusion
characteristics, stochastic distribution of SN in the Galaxy,
etc., are given in \cite{Svesh}.  We refer to the particles as `protons'
but the results will refer to nuclei of the same rigidity of another 
element, with a suitable change in ordinate to represent the abundance 
of that nucleus. The Figures show our estimate of the energy spectrum of
the proton component. In Figure 1a we show how the stochastic
distribution of SN modifies the shape of the CR energy spectrum
for just a single class of SN. The Figure shows 50 spectra
produced by standard SN with $logE_{51}$=0 and $logE_{max}$=5.6
distributed randomly in space and time in the Galactic Disk. Here
$E_{51}$ is the explosion energy in 10$^{51}$erg units. Figure 1b
shows the mean of these spectra and their standard deviation (we
refer to the standard deviation of the logarithm of the intensity
at a particular energy as the `irregularity' at that energy [5,
19]). In the real Galaxy, grouping of SN in both space and time
would cause the spread in spectra to be even bigger. Moving to
Figures 1c and 1d, these show 50 spectra and their mean,
respectively, for the case where $logE_{51}$ has a log-normal
distribution around $logE_{51}$=0 with a standard deviation of
$\sigma_{logE_{51}}$=1.2. $E_{max}$ has been taken dependent of
the explosion energy as $E_{max} \propto \sqrt{E_{51}}$
\cite{Svesh}.
\begin{figure}[htb!]
\begin{center}
\includegraphics[height=15cm,width=12cm,angle=-90]{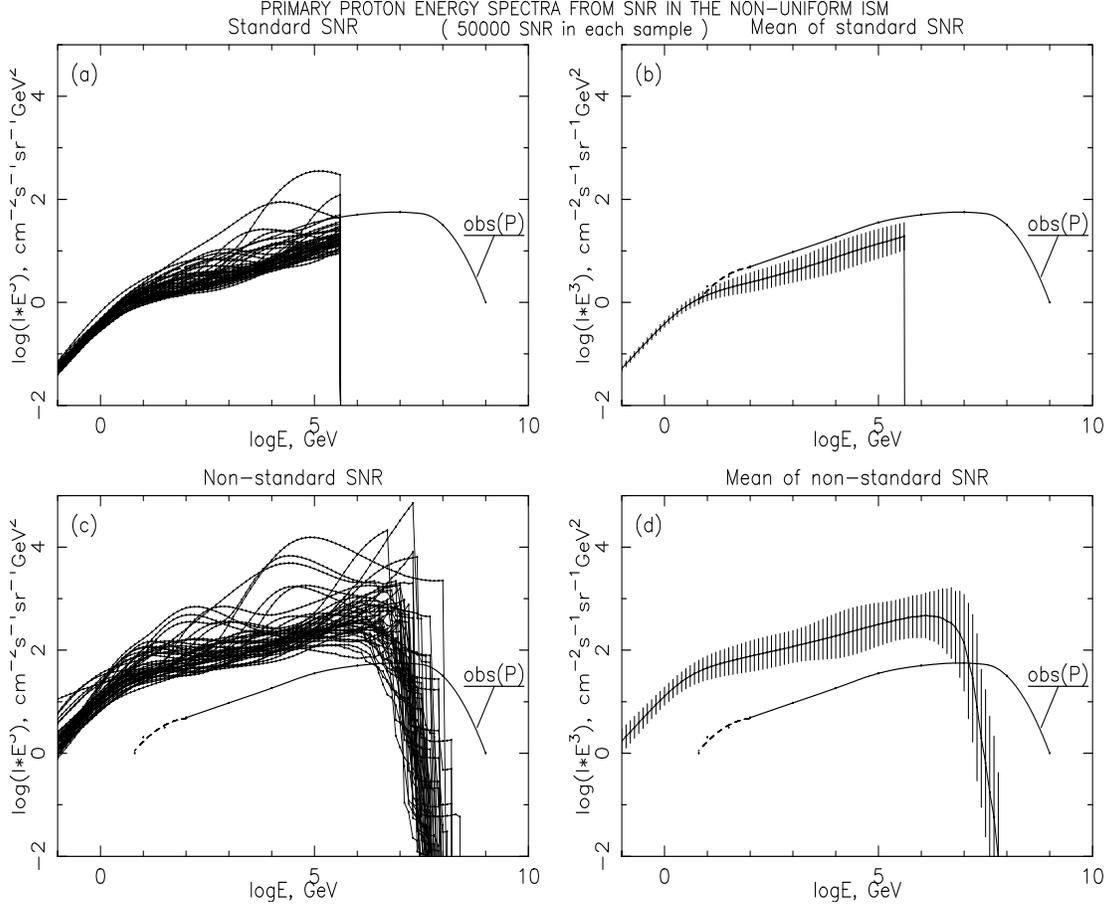}
\caption{\footnotesize Energy spectra of CR accelerated by SN (~50 samples with 50000
SN in each~): (a) with a standard
explosion energy $logE_{51}$=0 and a maximum particle energy $logE_{max}$=5.6;
(b) the mean spectrum and its irregularity for samples shown in (a) - the irregularity
is the standard deviation from the mean CR intensity for different samples of
the spectrum indicated by vertical lines; (c) with a
log-normal distribution of an explosion energy around $logE_{51}$=0 with a standard
deviation of $\sigma_{logE_{51}}$=1.2 and $E_{max} \propto \sqrt{E_{51}}$; (d) the mean
 spectrum and its irregularity for the samples shown in (c). The curve {\em obs(P)}
 is our estimate of the proton spectrum (~or the rigidity spectrum for heavier nuclei~)
 needed to fit the experimental data \cite{EWW}.  `E'$_{51}$ is the explosion energy in units of $10^{51}$ erg}
\end{center}
\label{fig:beyond1}
\end{figure}
It is seen that the spread of explosion energies leads to
\begin{itemize}
\item a displacement of the knee energy from $logE_{max}$=5.6 up
to 7-8 and makes the knee smoother \item an increase of the CR
intensity by up to an order of magnitude \item an increase in the
degree of irregularity of the spectrum.
\end{itemize}
The first two items are a consequence of the distribution of $E_{51}$ which is made
on a log-normal scale while the CR intensity is $\propto E_{51}$. The rise of the
irregularity is a consequence of the introduction of the big explosion energies which
 increase the fluctuations. Evidently a smaller value of  $\sigma_{logE_{51}}$
than adopted would give consistency between predicted and {\em 'obs(P)'}, but only at
sub-PeV energies.

Figure 2 shows the CR spectra produced by the two main types of SN
which are thought to contribute to
 the bulk of CR: SNIbc and SNIIn \cite{Svesh}. The explosion energy of both SN is the
same as in Figure 1 {\em ie} distributed around $logE_{51}$=0 with
$\sigma_{logE_{51}}$ =1.2. $E_{max}$ of SNIbc is also the same,
{\em ie} $\propto \sqrt{E_{51}}$, and distributed around
$logE_{max}$=5.6. $E_{max}$ of SNIIn, due to the different
progenitors and the environment, is 15 times higher and
distributed around $logE_{max}$=6.8. The fraction of SNIbc among
all SN is 0.2, that of SNIIn is 0.1 \cite{Svesh}. Figures 2a and
2b show
 50 samples of the total spectra produced by both SNIbc + SNIIn and their mean
respectively. Figures 2c and 2d show the contribution of only
SNIIn and their mean.
\begin{figure}[htb!]
\begin{center}
\includegraphics[height=15cm,width=12cm,angle=-90]{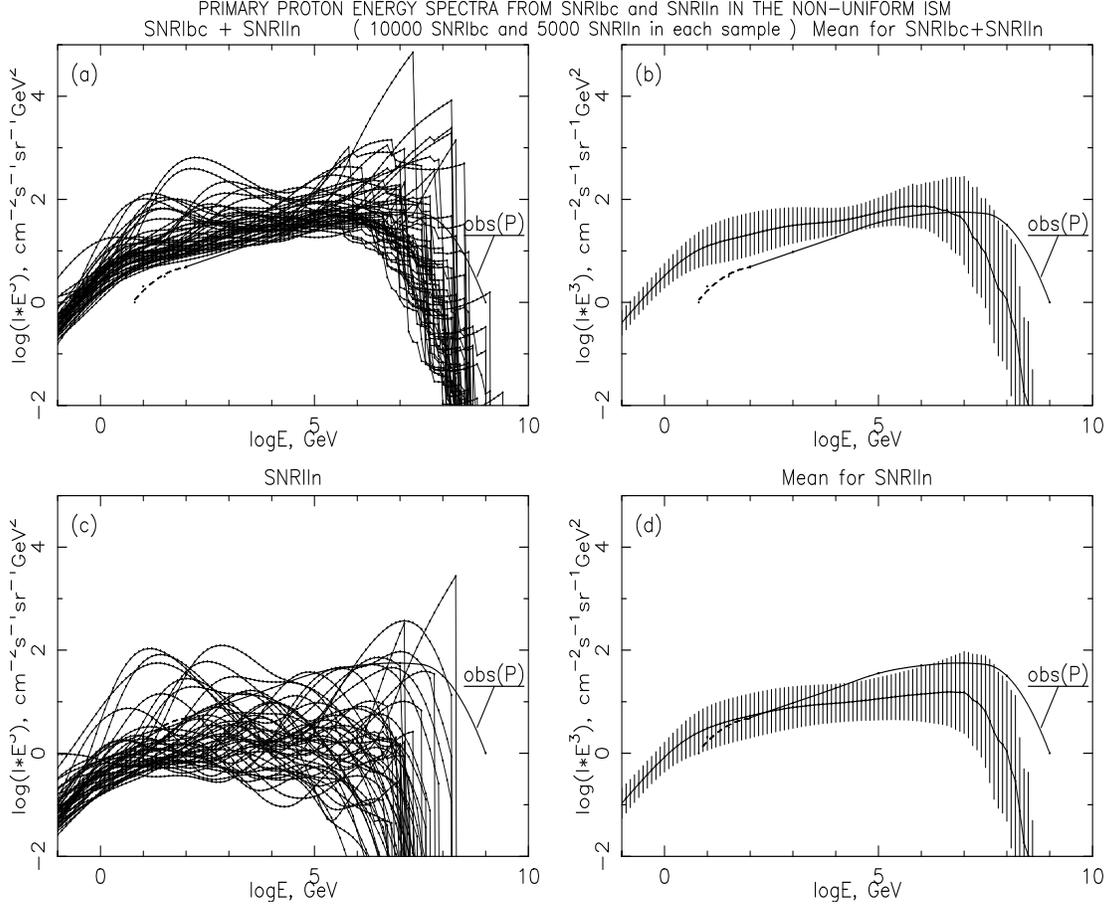}
\caption{\footnotesize Energy spectra of CR accelerated by two types of SN - SNIbc and
SNIIn (~50 samples with the same distribution of $logE_{51}$ for both, but different
distributions of $logE_{max}$ (~see text~)): (a) the summed spectra of 10000 SNIbc and
5000 SNIIn; (b) the mean spectrum and its irregularity for samples shown in (a);
(c) the contribution to the total spectrum of only SNIIn; (d) the mean
 spectrum and its irregularity for SNIIn samples shown in (c). As in Figure 1 the
irregularity in Figures 2b and 2d is the standard deviation from the mean CR intensity
for different samples of the spectrum indicated by vertical lines. The curve
{\em obs(P)} is as in Figure 1.}
\end{center}
\label{fig:beyond2}
\end{figure}

It is seen that
\begin{itemize}
\item the total CR intensity decreases by a factor of 3 compared
with Figure 1c and 1d due to the decreased SN rate (~30\%~) \item
the irregularity of the spectra increases because of the presence
of the rarer type of SN - SNIIn \item the maximum energy increases
by about an order of magnitude \item the sharpness of the knee
decreases due to the increase of $E_{max}$ (although it could, of
course, be restored by involving the presence of a close and
recent SN).
\end{itemize}

Figure 2 demonstrates that CR beyond the knee cannot be produced
just by the adopted fractions of SN with a higher $E_{max}$. The
distribution of the explosion energy and the maximum particle
energy combined with a reduced rate of explosions give rise to an
extreme irregularity of the expected spectra, which is in fact not
observed and there is still a high energy gap to be filled. The
remedy could lie in an occasional even stronger source; this
'stronger source' may be a yet more effective SNR using the model
of Bell \cite{Bell}. Here, SNR expanding into dense molecular
clouds, compress the magnetic fields and rapidly accelerate
particles to very high energies. We call these SNR, BSNR. The
ensuing spectra would be even more variable, however. A Giant
Galactic Halo which traps the particles is another possibility;
such a Halo could smooth out the BSNR fluctuations and it may, or
may not, provide further acceleration.

A mixed mass composition in the {\em energy} spectra can reduce
the irregularity of the spectra, since it displaces the same
rigidity spectra for different nuclei with respect to each other
when plotted in terms of energy, and the excursions in intensity
cease to be completely in phase. However, this effect is small
below the rigidity cutoff and only makes the cutoff in the energy
spectrum less sharp compared with the cutoff in the rigidity
spectrum (~see Appendix~).
\subsection{Pulsars}

\subsubsection{General Remarks}

In a recent paper \cite{EW7} we made the case for pulsars being serious candidates for
the origin of particles beyond the knee. Some general remarks are necessary first. When
 a different type of source is postulated for energies above the knee, the
(~comparatively~) smooth join of the two spectra is a worry. In the case of pulsars,
the near equality of the total kinetic energy carried by an SNR and the average initial
 rotation energy of a pulsar means that the join might be expected to be not too
traumatic.

\subsubsection{The CR Energy Spectrum from B0656+14}
In \cite{EW4,EW8} we claimed that the SNR Monogem Ring is a good candidate for such
 a  Single Source. It has been shown that the pulsar B0656+14, which is close to the
morphological center of the Monogem Ring, is associated with this SNR and this is a
potential further source of CR. The rotation period $P$ of this pulsar is
 0.3848 sec, the spin-down rate \.{P} is 5.5032$\cdot$10$^{-14}$. In the rotating
dipole model these parameters correspond to a very narrow peaked
emission of cosmic rays at a rigidity of 0.25 PV \cite{EW7}, a
rigidity remarkably close to the knee. This rigidity corresponds
to a knee energy of about 3 PeV for nuclei as heavy as Mg where it
would give a very narrow peak of its energy spectrum, and cannot
be invoked in its simple form.  However, the pulsar could
contribute to energies above the knee if cosmic rays emitted by it
in the past were confined within the SNR for a long time after its
explosion and released not long ago. If the pulsar emits
predominantly iron nuclei they can contribute to the cosmic ray
energy spectrum up to well above the knee.

Figure 3 shows the results of our calculations for MONOGEM \cite{EW7}
assuming the emission of just protons and iron nuclei and choosing
an appropriate trapping time.  The propagation parameters were our
usual ones \cite{EW3}.  Admittedly the model is rather ad hoc but it is
physically attractive.  However, the energetics are demanding.
`P' requires that most of the rate of loss of rotational energy of
the pulsar goes into protons, a more reasonable fraction (3$\%$)
is needed for iron nuclei.  A rather serious problem concerns the
expected anisotropy if most of the CR came from a single local
pulsar.

\subsubsection{High Energy CR From Many Pulsars}

A more likely scenario is that a number of pulsars contribute
perhaps most of the CR flux above the knee.  This model has its
adherents.  In one such model \cite{Gille} a distribution of pulsar
periods at birth is chosen - which is not unreasonable - and iron
nuclei are involved such as to give the measured CR spectrum.  The
energetics are reasonable.  The anisotropy problem is reduced by
postulating a Giant Halo (20-50 kpc radius), as is required in our
own analysis.

The problem encountered by all the pulsar models is the need for
young pulsars, with periods as low as a few ms; the presence of
such young pulsars in sufficient numbers is subject to debate but
progress is being made.

We conclude that pulsars can be considered as contenders for the
origin of CR above the knee if coupled with a Giant Halo.

\begin{figure}[htb!]
\begin{center}
\includegraphics[height=6cm,width=8cm]{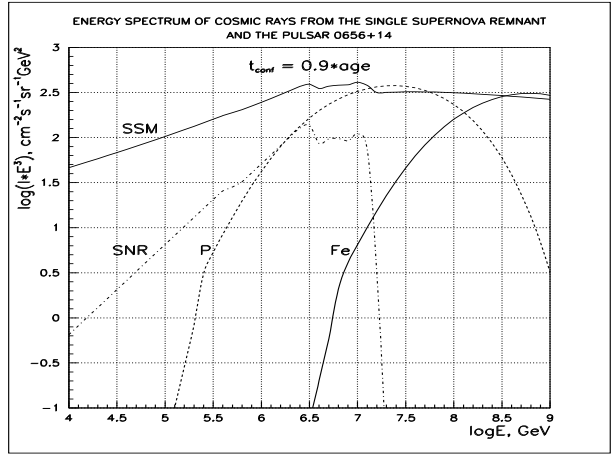}
\caption{\footnotesize Energy spectra of CR from the pulsar B0656+14, observed at the
Earth. They are calculated for a confinement time $t_{conf}$, during which cosmic rays
 were confined within the SNR shell, lasting 0.9 times the pulsar age. The CR nuclei
emitted by the pulsar are protons ($P$, dashed line) and iron-nuclei ($Fe$, full line),
 the efficiency of proton acceleration is near unity and that for iron acceleration is
 taken as 0.03. The CR energy spectrum in
the Single Source Model (which fits observations) is shown by the
full line denoted as 'SSM' with the contribution of the SNR shown
by the dash-dotted line denoted as 'SNR' }
\end{center}
\label{fig:beyond3}
\end{figure}

Such a scenario has a number of attractive features. The first is
that the mass composition above 40 PeV continues to be heavy due
to the dominance of iron nuclei stripped from the pulsar surface
up to a second knee at $\sim$400 PeV. The second is that the shape
of the energy spectrum is close to the observed one. Very young
and very rapidly spinning pulsars may have contributed to the
'gap' in Figures 1 and 2 (i.e. the range logE: 8-10) by way of
their particles being trapped in the Galactic Halo.

\subsubsection{Fine structure in the region of 100 PeV}

A topic which has relevance to CR production by pulsars,
particularly, because of the likelihood of a near monogetic beam
being generated by a pulsar of a particular age is the presence of
'fine structure' in the primary spectrum. Consequently, we have
searched for fine structure in the observed extensive air shower (~EAS~) 
size spectra in the region beyond the knee up to 1 EeV (~1000 PeV~) by
the same running mean method as we did in \cite{EW9}. In order to
enlarge the statistics of the data in the energy region above 100
PeV we added to the 40 EAS size spectra presented in earlier work
6 recent size spectra from the GRAPES
 III EAS array \cite{Tonw}. The mean excess of the running means is shown in Figure 4
together with that derived from recent MSU results on EAS muon
size spectra \cite{Fomin}.
\begin{figure}[htb!]
\begin{center}
\includegraphics[height=15cm,width=8cm,angle=-90]{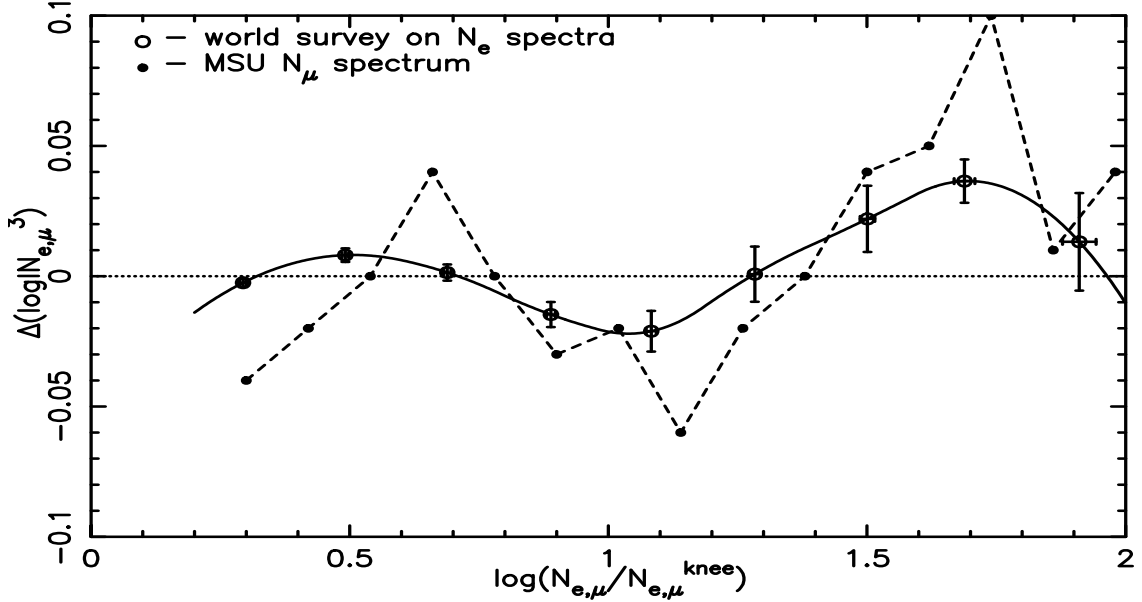}
\caption{\footnotesize The mean excess over the running means for the 46 EAS
size spectra (~full line~) and the MSU muon size spectrum \cite{Fomin}
(~dashed line~).}
\end{center}
\label{fig:beyond4}
\end{figure}
 The small excess at $log(N_{e,\mu}/N_{e,\mu}^{knee}) = 0.5$ corresponds to our second
peak in the primary energy spectrum at about 12 PeV. Here $N_{e,\mu}$ are total numbers
 of electrons or muons in the EAS and $N_{e,\mu}^{knee}$ are corresponding numbers at
the knee.  We associated the second peak with
 the cutoff of iron nuclei accelerated by the Single Source SNR. However, in the
region of $log(N_{e,\mu}/N_{e,\mu}^{knee}) = 1.3-2$ there is
another positive excess of running means with evidence for a
positive curvature of the spectrum at about 100 PeV. The magnitude 
of the excess at $\sim$100 GeV is significant in the EAS data ($\sim 4\sigma$)
and the significance is increased still further by the peak at the same energy 
in the muon data. The shape of
the iron spectrum from the pulsar B0656+14 shown in Figure 3 has a
similar positive curvature in the 100 GeV energy region, although it is
less sharp. It should be noted that the 'peak' at $\sim 100$ PeV
is at a higher energy than the 40 PeV we would expect if the knee
were due to helium rather than oxygen and the relevant CR were
accelerated by SNR.

The third advantage of the pulsar scenario is that the dominance
of just a single pulsar above the second (~iron~) peak of the SSM
gives a smooth spectrum with a small irregularity beyond the knee
which is the feature actually observed. It can be remarked that
the rigidity of iron nuclei at $logE = 8.5$ is probably still too
low for
 directional anisotropies associated with this nearby SNR (~Monogem~) to be detected
(~see \S 3.2.6~).

\subsubsection{The Confinement of CR Generated by Pulsars inside SNR}

In the case of pulsar acceleration of CR there is the need for
slow release of the particles.  This is where the associated SNR
should have relevance.  The interaction between a pulsar and its
surrounding nebula has been examined in many publications (~see \cite{Bedn} and
references therein~). In \cite{Bedn} the confinement of particles
 generated by the pulsar inside the nebula is limited by the time taken for its
 diffusion radius to reach the shell created by the SN shock wave. In the framework of
our model of the SN explosion \cite{EW9} this time $T_c$ is determined by the equation:
\begin{equation}
L(\frac{T_c}{\tau(E)+L/c})^{1/\alpha} = R_0(\frac{T_c}{T_0})^\beta
\end{equation}
where the left hand side is for the diffusion radius and the right hand side is for the
 radius of the shock
wave. Here $L$ is the characteristic distance reached by the
particles of energy $E$ in the process of diffusion during their
lifetime $\tau$ and $c$ is the speed of light, $R_0$, $T_0$ are parameters
which determine the speed of the SNR radial shell expansion. We introduce the
term $L/c$ to limit the speed of diffusion to the speed of light,
which is important at energies approaching 0.1 EeV. $\alpha$ is
the coefficient which determines the character of the diffusion:
$\alpha = 2$ for regular, 'gaussian', diffusion in a uniform
interstellar medium, and $\alpha = 1$ for anomalous diffusion in a
non-uniform turbulent medium with a Kraichnan-like spectrum of
turbulence \cite{Lagu,EW10}. On the right hand side of equation (1) we have 
the radius of the SNR shell as a function of time represented by 
$R(t) = R_0(\frac{t}{T_0})^{\beta}$ with $R_0 =
50 pc$, $T_0 = 2 \cdot 10^4$ years and $\beta = 0.5$. If we want
this equation to be fulfilled for our standard confinement time,
$T_c = 8 \cdot 10^4$ years, during which the radius of the shell
created by the shock wave reaches 100 pc, we have to assume that
the ordinary diffusion coefficient (~$\alpha = 2$~) for the
$\sim$100 PeV particles which we should like to confine inside the
shell, is $D \approx 2\cdot 10^{28}cm^2s^{-1}$. It is much less
than the $\sim 2\cdot 10^{31} cm^2s^{-1}$ adopted by us for
anomalous diffusion outside the shell. An equivalent mechanism is
to assume that the SNR shell is not transparent to the particles
until the remnant 'bursts' (~see later~).

All these estimates are very approximate and given just for an illustration of the
 conclusion that the diffusion of particles generated by the pulsar inside the SN shell
 has to be much slower than in the ISM outside the shell. Some support comes from the 
HESS observation of the relatively
young ( a few thousand years ) SNR G0.9+0.1 near the Galactic
Center \cite{HESS2}. They could not see the extension of the
source and concluded that {\em the gamma-ray emission appears to
originate in the plerionic core of the remnant, rather than the
shell}. It could mean that CR accelerated by the pulsar diffuse so
slowly that even after a few thousand years they have not reached
the shell.

When, after a long time, the diffusion front of CR approaches the
SN shell, the CR particles do not necessarily escape. In our case
of particles emitted by a pulsar they catch up a receding shell
from behind, i.e. from its downstream region. They lose a part of
their energy accelerating and modifying the shock wave and appear
with a smaller energy within the shock. The loss of energy is even
larger if
 the particles are reflected back from the shock by a receding magnetic mirror.
Calculations show that the probability of such reflection is quite high \cite{Blasi}.
Thus, the scenario looks as if for the ambient and upstream CR particles the shock wave
and expanding SNR shell serve as an accelerator while for downstream CR they
serve as a trap.

If there is  a supersonic particle wind from the pulsar, driven by its rotation, then
one can expect the existence of a termination shock where its speed decreases enough
 to become subsonic and particles accelerated by the pulsar can be trapped inside
the smaller volume of the pulsar wind `nebula' limited by the
termination shock and their confinement time can be even longer
than for trapping in the associated SNR.

\subsubsection{The Anisotropy}

    The dominance of a Single Source such as the Monogem Ring SNR and its associated
PSR B0656+14 in the knee region and beyond can create problems
with the anisotropy of CR. However, there are a few indications
that the expected anisotropy does, in fact, exist. The global
search for an anisotropy, made in \cite{WW1,Clay}, has shown that
at primary energies $10^{16}-10^{17}$ eV there is an excess of the
CR intensity in the general direction of
 the Monogem Ring. More recent studies at lower energies \cite{Kuli,Benk} have also
indicated regions of excess intensity within the Monogem Ring, though at a low
confidence level. Therefore, we cannot exclude the possibility that this region of
the sky gives an enhanced intensity of CR in a wide energy interval from sub-PeV up to
tens of PeV energies.

\subsection{A Giant Galactic Halo}

\subsubsection{General Remarks}

If CR in the PeV - EeV energy range are of Galactic origin, with very energetic SNR and
 young pulsars as sources of their energy, then there needs to be a mechanism which
reduces the irregularity of their 'explosions' and the effect of the short time when
these sources are able to produce particles of the highest energy. The most natural way
 is to introduce a long term accumulation of CR within a confinement volume, similar to
 the well known 'leaky box' for the Galaxy, but with a much larger size and a longer
lifetime in it. Such accumulation can be achieved with the help of
a Giant Galactic Halo, a proposal that is not new but has
important implications \cite{Ginz,WW2,Voelk}.

In conventional models of CR origin and propagation it is assumed
that there is free escape from the Galactic Disk (~with a typical
scale height of $\sim$1 kpc~) into a Halo with, perhaps, a scale
height of $\sim$10 kpc. In fact, however, it seems likely that
there is a Giant Halo, extending to perhaps 100kpc (~or even
further~) from the Galactic Plane, in which there is plasma. Some
evidence has come from gas displacements in the Magellanic Clouds
which are interpreted as being due to the ram-pressure effect of
Galactic gas rotating with the Galaxy \cite{Moore}. Similar
arguments favoring the existence of big halos for galaxies and
even for clusters of galaxies have been put forward also in
\cite{Zirak,Suto,Veil}. In \cite{EW10} we showed that anomalous diffusion
can help to create such a Giant Halo with a long tail of CR above
and below the Galactic Disk.

A Halo, involving only diffusion, would not have a trapping effect as
such, however. CR would
simply diffuse out of the Disk due to the CR density gradient and not be accumulated.
One has to introduce a 'membrane' or 'reflector' to stop further leakage, to reflect
the outflow and to start the accumulation. It is here where the Galactic wind can help.
 Such a wind is to some extent similar to the Solar wind. It is due to the
gas and CR pressure gradient in the Disk, its starting speed is supersonic and it
 itself may well accelerate particles in the Halo if the Mach number is high enough.
Eventually, this wind loses its energy due to friction with the ambient gas in the
 Intergalactic Medium (~IGM~) and at some distance from the Disk its speed becomes
subsonic. This is the region where the 'termination shock' is formed. This termination
shock can act to reflect and trap CR particles and perhaps accelerate them
\cite{Voelk,Joki1,Joki2}.

Table 1 shows characteristics of a possible Giant Halo with (near-)reflecting walls,
largely using results derived by us from the data given in \cite{Voelk}. The figures
given allow the energetics of the phenomenon to be put in perspective. The next
subsection provides more details.

It must be said that there is no direct experimental evidence that the Galactic wind
and its termination shock exist. They have been proposed as an analogy of the solar
wind, which is detected by various spacecrafts and where there are indications from the
 Voyager I mission that the spacecraft is approaching the termination shock region in
the heliosphere \cite{Stone}. However, the
assumption about the Galactic wind and the termination shock helps to fill the gap
between the Galactic CR at 40-50 PeV and EG CR which may appear beyond the 'ankle'
at a 3-8 EeV \cite{Wibig}. There is an alternative model, however, in which this
interval is filled by the true EG CR and the transition occurs at a lower energy
\cite{Bere1}. The assumption also helps us to understand a number of observed CR
phenomena viz.
\begin{itemize}
\item the small radial gradient of CR intensity \cite{Breit}
\item the small irregularity of the CR energy spectrum (~quantified in the Appendix~)
\item the small anisotropy of arrival directions.
\end{itemize}

The main aspects of the Giant Galactic Halo will be considered below. We discuss Halo
acceleration, diffusion and simple trapping.

\subsubsection{Halo Acceleration}

In \S 3.1 we showed that the admission of a variety of explosion energies as well
as a variety of SN types with some of them having much higher maximum energies
$E_{max}$ for the acceleration (~'hypernova'~) allowed one to get CR particles above
 PeV energies (~although, in its considered form, the maximum energy was
 still not high enough and BSNR and/or young pulsars were needed~). However, the paucity of such sources and
 the high speed
of the diffusion results in a very big irregularity of CR energy spectra in this
region. Even if one restricts the speed of the diffusion to the speed of light, as one
must (~formula 1~) it does not reduce the irregularity very much (~Figure 1~).
Therefore, we give
preference to a more regular mechanism where high energy particles appear as the result
 of re-acceleration of lower energy CR particles emitted by the Disk by the shock waves
 in the Halo. The energy spectrum of these particles should be less irregular since the
 re-acceleration process seizes a large space in the Halo and integrates the spatial
irregularities \cite{Voelk} arising in the Disk; specifically, estimates indicate that
the anisotropies should be limited to a few percent (~see the Appendix~) and the
corresponding distortion of the spectral shape should be small. For the re-acceleration
 another source of energy supply has to be invoked in addition to the 'direct' CR
energy from SN explosions (~{\em eg}. Galactic wind, rotation of the Galaxy, the
remaining non CR - SNR energy, etc.~)

In what appears to have been the first attempt at making calculations for acceleration
in the Halo and at the termination shock \cite{Joki1,Joki2} attention was devoted to
detailed calculations of the maximum energy achievable. The authors
point out that the important acceleration time in a medium where the shock velocity is
 $V_{sh}$ and the diffusion coefficient is $D$ is $t_{acc} = 4D/V_{sh}^2$. Typical
shock velocities of $\sim 500 kms^{-1}$ are invoked and the result
is that, for a derived magnetic field of $\sim 0.07 \mu G$ at the
termination shock at a radius of 100 kpc, protons can be
accelerated to $\sim 6\cdot 10^{18}$eV (~i.e a rigidity of $6\cdot
10^{18}$V~). The diffusion coefficient adopted is presumably
$7.5 \cdot 10^{33}~cm^{2}~s^{-1}$.  We ourselves prefer a reversal
length of $\sim$ 10kpc \cite{EWW}, i.e. $D\simeq3\cdot 10^{32}cm^{2}~s^{-1}$,
for which the maximum rigidity would be lower: $3\cdot 10^{17}V$, i.e.
approaching $10^{19}eV$ for iron; such an energy is adequate for
Galactic CR.

Reverting to the earlier model it is interesting to note that the
field compares well with our $0.1 \mu G$ averaged over the Halo
and derived from an estimate of gas temperature (~$\sim 10^6K~$)
and density (~$10^{-3}cm^{-3}$~) \cite{EWW}. Specifically, the
shocked Halo will have a field of $\sim 0.1 \mu G$ at a radial
distance of 1/3 of the termination shock radius.

If for the moment we ignore the need for an additional source of
energy for the re-acceleration and take the CR injection from the
Disk into the Halo as the only source of energy, its luminosity is
about $5\cdot 10^{40}erg \cdot s^{-1}$  from our model of a SN
explosion [25]. With a radius to the termination shock of 100 kpc
and an accumulation time of $10^{10}$y (~the age of the Galaxy~)
the corresponding energy density in the Halo is about $0.1
eVcm^{-3}$. The energy density of {\it observed} CR with energy
above 40 PeV is about $4\cdot 10^{-6}eVcm^{-3}$. Therefore, if we
associate CR above 40 PeV with CR from the Halo they should
contribute
 also below 40 PeV, even for the case of the minimum energy supply. The spectrum of
Halo CR at lower energies should be flatter than that with $\gamma = 3$ so as not to
contradict observations (~a retarding effect of the Galactic wind is postulated. In
this case, the Galactic sources would be spasmodic, with no nearby ones at the
moment~).

In the SSM we composed the CR energy spectrum from two components: the spectrum of the
Single Source and a so called 'background', the source of which we did not specify, but
 argued that it was created by the bulk of older and remote SNR, at least at energies
below the knee. If we calculate the energy density created by this
background we find that the minimum value of $0.1 eVcm^{-3}$
corresponds to CR above $\sim 10^2 GeV$, which could be the 'low
energy' cutoff for CR reaccelerated in the Halo. The real spectrum
of CR particles in the Halo should extend to lower energies than
$\sim 10^2 GeV$, particularly if an additional source of energy is
invoked in the Halo.

Turning to the irregularities, due to the integration of
individual spectra from different SN over the long accumulation
time and large volume, the irregularities created by the
stochastic nature of SN explosions will be smoothed and the
resulting irregularity of the CR energy spectrum in the Halo
should be close to zero. We shall demonstrate this fact later in
\S 4 and the Appendix.

A more recent and more comprehensive model with acceleration at
the termination shock has been given in \cite{Voelk}. The salient
points are given in Table 1. Prominent features are that it is the
pre-existing Galactic CR that are accelerated in the Halo, and few
of the post-1PeV particles are able to get back into the Galactic
disk. It will be noted that the required energy density of CR
above 1 PeV can be realised and that the mechanism stands the test
of energy conservation, viz. sufficient energy is injected from
the Galactic Wind and from Galactic rotation to produce what is
needed. (~see Table 1~).

There are other models proposed to accelerate CR particles up to EeV energies. Among
them is the 'cannonball model' \cite{Plaga,Dar,DeRuj} where the sources of energy is
the same
SN explosions but the accelerators are not only their shock waves, but predominantly
relativistic plasmoids emitted by SN into the Halo. Whether this mechanism
 is valid or another, such as the acceleration by multiple shocks \cite{Bykov}, it is
desirable that a substantial contribution from the Halo be provided even at lower
energies to reduce the
irregularity of the CR energy spectrum compared with the case where only SNR from the
Galactic Disk are contributors and there is no further accumulation outside the Disk.

Some other relevant points can now be mentioned.

(i) There is the well-known shear in the rotation of the Halo as a function of height
above the Galactic Plane. The moment of inertia of gas in the Halo is of the order of
$10^{62}erg$; this can be compared with a CR energy input from the Disk in $10^{10}$y
of $\sim 10^{58}erg$ - thus if only a small quantity of the shear energy finds its way
into CR the energy gain can be large, particularly at very high energies where the
particles can reach the termination shock. Here, where the magnetic field is 'wound
up' or compressed by shocks, there are ideal ingredients for the acceleration of very high energy CR.

In fact the transport of CR in a rotating scattering medium has
been put forward \cite{Webb} as a way of `energizing particles' but no
details appear to have been given.

(ii) With a Giant Halo of the type proposed, the vast majority of CR electrons will be
 absorbed - the so-called EG radio background will then have a big component from the
Halo.

(iii) Effects on the CMB might be expected via the Sunyaev-Zeldovich effect, similar to
 those detected from galaxy clusters \cite{Myer}.

Other sources of energy relevant to the Galactic Halo model - in
addition to very rare BSNR and millisecond pulsars - are as
follows:

(i) A non-negligible amount of energy has been brought into the
Galaxy by way of the impact of dwarf galaxies and the energy in
the associated shocks will have accelerated CR, some of which may
still be trapped.

An example will suffice.  A `small' galaxy of mass
$10^{9}M_{\odot}$ colliding with velocity 400 km s$^{-1}$ will
bring in a kinetic energy of $1.6\cdot 10^{59}~erg$.  One might
visualize 100 such small galaxies arriving over the life of the
Galaxy, corresponding to $1.6\cdot 10^{59}~erg$.  As a datum, SN at
$10^{51}~erg$ each with a rate of $10^{-2}y^{-1}$ would
deliver a similar amount of `mechanical' energy.  Of course, for
conventional SN, where the CR energies are mainly sub-PeV, the
efficiency of energy transfer to CR is probably much higher for
SNR (a figure of 10$\%$ is often quoted).  However, in the
important post-PeV region, where sources are so difficult to find
(but the total energy content is smaller) galaxy collisions might
be important.

(ii) Past explosions in the Galactic Centre are similarly `in with
a chance'.  Interestingly, the energetics are rather similar.  If
(and it is a big `if') the Galaxy has undergone Seyfert-type
activity, perhaps due to the properties of the central black hole,
with a lifetime $10^{8}y$ and a luminosity of $10^{10}L_{\odot}$,
 commonly quoted, then, for 10 such explosions in the lifetime of the Galaxy,
we have $6\cdot 10^{58}$ erg released.  The values are chosen so that
the Galaxy would appear as a Seyfert some 10$\%$ of the time - the
Universal average.

\subsubsection{Halo Diffusion}

Some remarks are now necessary about CR diffusion in the Halo.

Although the 'magnetic properties' of the Giant Halo are not known
with any accuracy some rather general estimates have been made in
\cite{Voelk} and in \cite{EWW}. In the latter, the mean field
$B$ and reversal length $\lambda$ were taken to be such that
$B\lambda$ changes only slowly with distance from the Plane. In
the Galactic Plane the mean field is $B \sim 2 \mu G$ and the
reversal length $\lambda \sim 0.2 kpc$; at a height of 100 kpc the
corresponding mean field is $B \simeq 0.1 \mu G$ and $\lambda
\simeq 10 kpc$.

The fall off of field with height can be understood in terms of
the expanding field-carrying SNR shells expanding out into the
Halo; the same phenomena gives the increased reversal length with
increasing height.  The fall in plasma density with height can
also be involved if, as in commonly assumed, there is a measure of
equality in energy density between magnetic field and plasma
energy.

It is appreciated that the above is an oversimplification because what is important for
 particle propagation in the Halo is the power spectrum of the irregularities - the
re-entrance of a particle being influenced largely by the power contained in lengths
of the order of the Larmor radius. For
$<B \lambda>$ $\sim 0.3 \mu Gkpc$, taking $\lambda$ as the Larmor
radius, the corresponding energy - for protons - is given by $logE = 8.5$. In view of
$\lambda$ being identified with the length scale having most power, the above means
 that there should be efficient scattering, and thus a degree of trapping, up to this
energy. It can be remarked that $logE = 8.5$ for protons is an important energy because
 applied to iron nuclei of the same rigidity the energy would be $logE(Fe) \simeq 10$
i.e. the likely energy limit for Galactic iron nuclei.  The spectrum denoted as
'obs(P)' in Figure 1 shows this is just where the measured proton spectrum begins to
fall.

In the original work \cite{WW2} it was pointed out that if the highest energy particles
 are trapped for a very long period by the termination shock, some nuclei will fragment
 and re-entrant protons will be mistaken for EG particles. Similarly, many apparently
EG particles will be trapped Galactic heavy nuclei.

Although the energetics of the Halo acceleration model appear to be acceptable, we know
 neither the expected spectral shape nor the manner in which the lifetime in the Halo
depends on energy. Thus, a rigorous derivation of the Halo spectrum cannot be given.
Nevertheless we give below in Figures 5,6 and 7, Halo spectra which satisfy the
boundary conditions.

\subsubsection{Halo Trapping, without Acceleration}

Following the 'General Remarks' in \S 3.3.1 we can examine the
Halo trapping model in more detail. This alternative scenario is
to postulate an emission spectrum from the Galactic Disk into a
Giant Halo which has a termination shock which acts as an
energy-dependent reflector but without acceleration. The emitted
spectrum would have a longer 'tail' than discussed earlier and
could include CR from very rare SNR in which they expand into
molecular clouds and compress the magnetic fields considerably,
i.e. BSNR \cite{Bell} and pulsars at very early
 stages. (~see \S 3.2~). The essential point here is that propagation in the
Galactic Disk will be so rapid that the chance of our seeing a
very energetic source at any one instant
 is very small; the particles will be trapped in the Giant Halo, however, and, smoothed
 out in space and time, will contribute to the ambient CR flux at earth (~see
Appendix~). We adopted a similar philosophy \cite{EW1} to account for EG particles
having come from galaxies basically similar to our own.

The lifetime inside the trapping Giant Halo can be much longer than without trapping
(~eg. $\sim 10^{10}$y for PeV protons~) and the expected CR intensity higher; more
particularly, the energy requirements for
Galactic sources can be less severe.

1. Galactic Centre explosions. If there is indeed a Giant Halo
that can trap particles for very long times, then the postulated
explosions in the Galactic Centre \cite{Said} the last being perhaps
$10^8$ years ago - can have populated the Galaxy with 'low' energy
CR. Certainly, one would expect the low energy particles to be
trapped for the longest period of time - and thus the spectrum to
be steep. The energy release referred to earlier ($6\cdot 
10^{58}~erg$. It must be admitted that the efficiency of conversion
into cosmic rays would have to be high even if one postulates a
falling CR density with distance from the Galactic Plane.

2. SNR in the Giant Molecular Ring at $R\sim 3kpc$ where the ambient ISM gas density is
 high may have produced many low energy CR and, again, these will be resident in the
Halo. Concerning the overall energetics in the Halo the conventional sources of energy
in the Disk are adequate (~Table 1~), provided that the Halo trapping time is
essentially the age of the Galaxy.

It is necessary that the Galactic emission spectrum extends beyond the end of the 'two
types of SN' of Figure 2, because of the presence of BSN.
We write the Galactic emission spectrum as $I_{em}^g(E)$.

If there is trapping during the time $\tau$ which depends on
energy (rigidity) as $\tau(E/Z) = \tau_0 \cdot (E/Z)^{-\delta}$ (E
being the energy in GeV and $\tau_0$ being the mean lifetime
at 1GeV) then in the framework of the Giant Halo model the energy
spectrum of CR accumulated during the accumulation time $T$ is
like that in a 'Giant Leaky Box':
\begin{equation}
I_{Halo}(E)=I_{em}^g(E) \int_0^T \frac{V_{SNR}}{V_{halo}} \nu exp(-\frac{T-t}{Z^\delta \tau_0 E^{-\delta}})dt
\end{equation}
Here $V_{SNR}$ and $V_{halo}$ are volumes of the SNR shell and the Halo, the ratio of
which in the case of spherical symmetry can be taken as the cubed ratio of radii
$R_{sh}$ and  $R_{halo}$, where $R_{sh}$ is the radius of the shell at the moment when
CR are deconfined and begin to diffuse in the ISM, $\nu$ is the SNR explosion rate and
$t$ is the time. If there is no evolution, and all the variables
under the integral do not depend on time or on coordinates inside the Halo, the
integration is straightforward and gives the result:
\begin{equation}
I_{Halo}(E)= I_{em}^g(E) (\frac{R_{sh}}{R_{halo}})^3 \frac{\nu Z^\delta \tau_0}{E^\delta}[1-exp(-\frac{TE^\delta}{Z^\delta \tau_0})]
\end{equation}
It can be seen that at low energies the energy spectrum in the Halo follows the
emission spectrum, at high energies its slope index is higher by the factor 
$\delta$. The absolute intensity differs from the emitted intensity
by the 'multiplication' factor which is simply $\nu T$ at low energies (~Figure 5a~)
and a 'dilution' factor of $(\frac{R_{sh}}{R_{halo}})^3$. The steepening of the
spectrum occurs at the energy $E = \frac{\tau_0}{T}$ GeV.

As an example, we present in Figure 5b the energy spectrum of protons (~ie the rigidity
 spectrum of all nuclei~) in the Halo, adopting an emission spectrum
$I_{em}(E) = AE^{-2.7}$ valid at all energies, $R_{sh}=0.1$kpc
(~our model value \cite{EW9}), $R_{halo}=100$kpc, $\delta = 1$
which corresponds better to ISM conditions at the Halo boundary
[55], $T=10^{10}$years - the age of our Galaxy and
$\tau_0=10^{17}$years. This 'lifetime' is clearly higher than the
age of the Galaxy, but it simply means that low energy CR never
leave the Halo and do not populate EG space. In Figure 5b we show
also the so called EG energy spectrum, which is simply related to
the spectrum of particles leaking from the Halo into outer space.
We shall discuss it later.
\begin{figure}[htb!]
\begin{center}
\includegraphics[height=12cm,width=15cm]{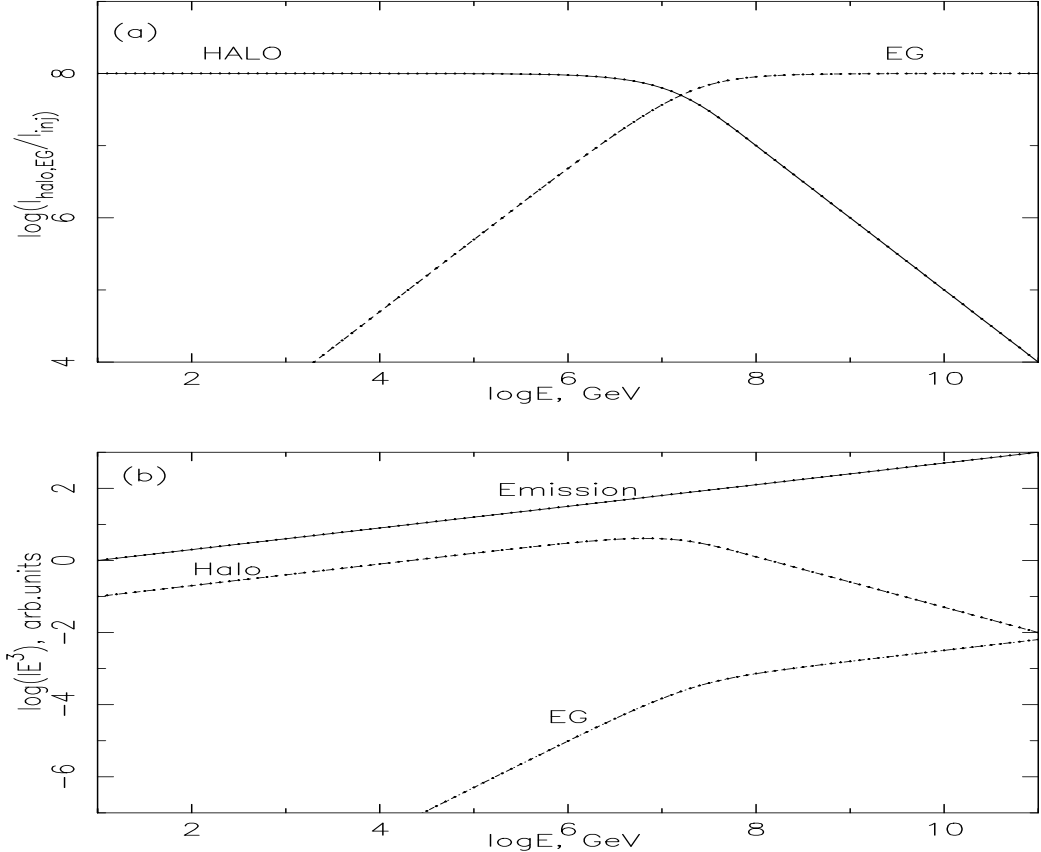}
\caption{\footnotesize (a) Multiplication factor: the ratio of
Halo (~full line~) or EG (~dashed line~) spectra to the emission
spectrum with no spatial dilution. (b) An example of emitted Halo
and EG spectra with multiplication and dilution for protons and
parameters described in the text.}
\end{center}
\label{fig:beyond5}
\end{figure}

\subsection{Extragalactic Cosmic Rays}

A possible scenario is that leakage from the Halo of galaxies {\em similar to our own}
is the source of CR populating EG space. In this case we can write the equation:
\begin{equation}
I_{EG}(E)=I_{em}^g(E) \int_0^T V_{SNR} \rho_g \nu [1-exp(-\frac{T-t}{Z^\delta \tau_0 E^{-\delta}})]dt
\end{equation}
Here the only new notation is $\rho_g$ which is the spatial density of galaxies in
EG space. Assuming no cosmological evolution, we obtain
\begin{equation}
I_{EG}(E)= I_{em}^g(E) \frac{4}{3}\pi R_{sh}^3 \rho_g \nu \{T-\frac{Z^\delta \tau_0}{E^\delta}[1-exp(-\frac{TE^\delta}{Z^\delta \tau_0})]\}
\end{equation}
The multiplication factor and the EG spectrum corresponding to
$\rho_g = 6.4\cdot 10^{-2} Mpc^{-3}$ \cite{EWW} is shown in Figure 5. It
is seen that the contribution of EG CR is small for the chosen set
of parameters and cannot reduce irregularities in the spectrum
caused by the stochastic nature of SN explosions. Another caveat
is seen above EeV energies: the shape of the energy spectrum does
not agree with the experimental observations of the 'ankle'
\cite{Nagan}.

However, if we give up the assumption that all the galaxies are
identical to our Galaxy and admit that there are sufficiently
numerous much more powerful galaxies in the Universe, then the
possibility of agreement with experiment appears. The additional
assumption that CR emitted into the Halo have a mixed mass
composition improves this agreement. Figure 6 shows the Emission,
Halo, EG and total spectra for the 4-component mass composition
close to that indicated in \cite{Bier2} for 1 TeV per nucleus:
0.487-P, 0.310-He, 0.121-O and 0.082-Fe and for the 10-fold
reduction of the mean distance between EG sources. It is seen 
that all the main features of the all particle spectrum in the PeV-EeV
region and above are fairly well reproduced.
\begin{figure}[htb!]
\begin{center}
\includegraphics[height=15cm,width=9cm,angle=-90]{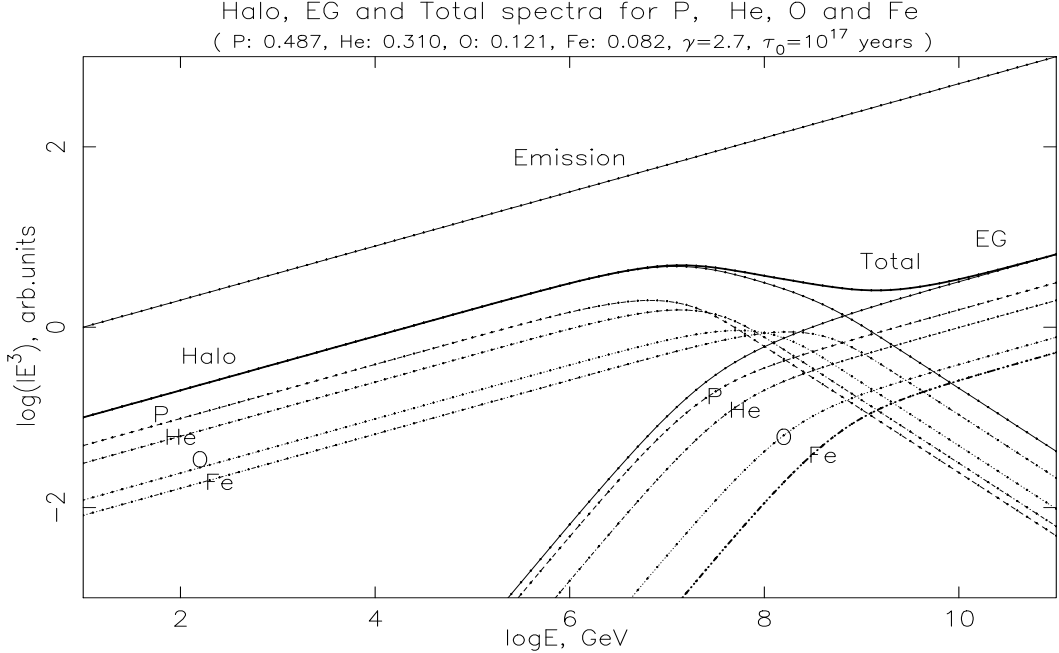}
\caption{\footnotesize Emission, Halo, EG and total spectra for the 4-component mass
composition: 0.487-P, 0.310-He, 0.121-O, 0.082-Fe. The EG mass composition at high
energies is the same as that in the Halo at low energy in this model. In practice the
EG composition and spectrum will be modified by propagation effects.}
\end{center}
\label{fig:beyond6}
\end{figure}
\section{Discussion}
\subsection{The preferred model}
We mentioned earlier that in our Single Source Model of the knee
we introduced the so called 'background' spectrum, which we
attributed to the contribution from many old and distant SN
\cite{EW5}. We now go further and assume that this contribution is
integrated in space and time within the Giant Halo and associate
this background with the model of the Halo + EG spectrum described
above. Figure 7a shows all four major components of the model: SNR
of the Disk, Halo, EG and SS. CR energy spectra from SNR were made
from standard rigidity spectra (~Figure 1a~) adopting the mixed
composition \cite{Bier2} at 1 TeV/nucleus. The total Halo + EG
spectrum of our model has been normalized to our phenomenological
background at 1 EeV. Figure 7b shows the sum of all four
components as the total spectrum with its irregularity. It is seen
that, due to smoothing effect of the Halo, irregularities below
the knee are reduced to the level observed in the experiment; the
standard deviation for $logE = 2.5 - 4.5$ is $\Upsilon = 0.07$ to
be compared with the experimental value of $\lapproxeq 0.05$ (~see
\S 3.3.2 and the Appendix~).
\begin{figure}[htb!]
\begin{center}
\includegraphics[height=12cm,width=15cm]{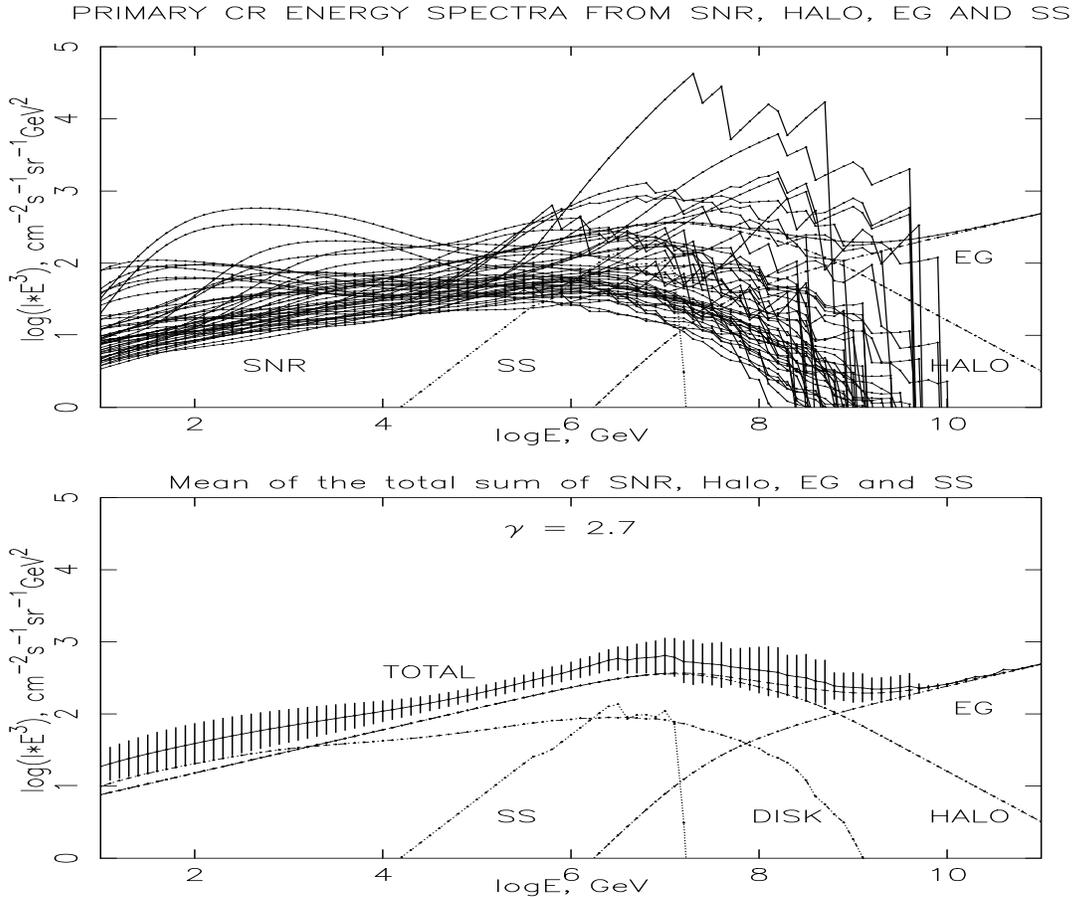}
\caption{\footnotesize (a) Four major components of the CR spectrum: SNR in the
Galactic Disk, Halo, EG and SS. (b) The all particle CR spectrum as the sum of four
components, shown in (a). The irregularity of the spectrum is substantially reduced
due to the smoothing effect of the Halo. The slope index of the emission spectrum in
the Halo is $\gamma = 2.7$.}
\end{center}
\label{fig:beyond7}
\end{figure}

The composition of the CR energy spectrum in our model is similar
to that in the so called 'three sites model' of Biermann [57]. The
difference is that we introduce a Giant Halo with extended
trapping as the smoothing mechanism, which makes
 the spectrum looking as a nearly regular power law.

We should like to underline the expense by which we achieved the general agreement
between our model and the experiment. It is:

(i) Introduction of a hypothetical Giant Halo as an intermediate
structure between the Galactic Disk and EG space. 

(ii) Assumptions about the Galactic Wind which creates the Termination
Shock at the boundary of the Giant Halo. The leakage from this
boundary makes the Halo similar to the Giant Leaky Box.
 
(iii) The emission spectrum of CR in the Halo has a power law shape with
slope index $\gamma = 2.7$ in the entire energy range from GeV to
hundreds of EeV. This value of $\gamma$ is essentially ad hoc but
there are, perhaps, ways of generating it, for example, explosions
 at the Galactic Centre, or the collisions of small galaxies.

(iv) EG CR appear as the result of the leakage from halos of all galaxies and some of
them should be much more powerful than our Galaxy.

There are attractive consequences of this model which agree with the observations
(consequences which were not put in at the start):\\
(i) the mass composition beyond the knee becomes heavier than below;\\
(ii) the EG CR also have a mixed mass composition with a predominance of protons,
because they escape from halos first (~propagation effects in EG space may reverse this
 result, however \cite{WgW}~);\\
(iii) the problem of the small radial gradient of CR is reduced due to the big
contribution of the Halo with its much more uniform spatial distribution of CR than in
the Disk;\\
(iv) the CR isotropy is explained by the same uniform spatial distribution in the Halo.

As for the model with no Halo, where the transition from Galactic to EG CR occurs
immediately after the termination of the former at $E_{max} \approx 0.04-0.1 EeV$
(~Figures 1, 2~) and at lower energies only SNR contribute with no substantial
EG fraction \cite{Bere1,Bere2}, then the problem of the big irregularity of the spectra
 expected in this
model remains unsolved. In this paper we do not examine the problems of the ultrahigh
energy, tens of EeV EG particles and their interaction with the cosmic microwave
background, nor the possibility that there are no truly extragalactic particles but
simply very energetic nuclei trapped in the Giant Halo \cite{WW2}.

Certainly, our model raises more questions than it gives answers.
At the present time it is purely phenomenological and some of its problems are:

(i) We are not sure which of the mechanisms give the particles above $E_{max}$ for SNR.
 It can be young supernovae (BSNR), young pulsars, cannonballs, re-acceleration in the
Halo or others. At the moment we prefer the re-acceleration mechanism because it seems
to be more regular than the others and works continuously, unlike pulsars or SNR.

(ii) We do not know why the so called emission spectrum in the
Halo needed to fit the experimental data, should have $\gamma =
2.7$, which is coincident with the spectrum in the Galactic Disk
but does not coincide with the slope of injection
 spectrum from our SN, which has the slope $\gamma = 2.15$; our suggestions of Galactic
Centre explosions and galaxy collisions, although reasonable, have
no possibility of proof.

(iii) It should be remarked that the Galactic Wind will affect propagation pf low 
energy particles generated in Disk sources and propagating therein. No allowance has 
been made for this effect.

(iv) The contribution of the Halo at GeV energies has to be reduced compared with that
 expected from the spectrum with $\gamma = 2.7$, at least at distances more than
50 kpc from our Galaxy, in order not to conflict with the low flux of $>0.1 GeV$
gamma-rays from Large and Small Magellanic Clouds \cite{Chi}; the clouds are inside the
 Halo
according to parameters adopted in the present model. It can be remarked, however, that
 in the work on acceleration in the Giant Halo \cite{Voelk} there is a radial gradient
which reduces the intensity at the  Magellanic Clouds sufficiently.

We hope to clarify some of these problems in subsequent papers.

\subsection{The conventional - but not preferred - situation}

In view of the fact that, conventionally, we expect the spectrum of particles escaping
from the Disk to have an index $\gamma = 2.15$, we have made calculations also for this
 situation, for the sake of completeness.

In Figure 8 we show the all particle spectrum as the sum of the four components: DISK,
HALO, EG and SS similar to that shown in Figure 7b, but for the case of the emission
spectrum in the Halo, having slope index $\gamma = 2.15$. The EG contribution is
increased by a factor of 8. The contribution of the Halo to the total CR intensity at
 energies below $10^4$GeV is negligible and the irregularity of the spectrum is
rather large (~that is why we prefer the steeper emission spectrum with $\gamma = 2.7$~).
Specifically, for the range $logE = 2.5 - 4.5$ the standard deviation of the sagitta is
$\sigma \sim 0.16$, whereas we find, experimentally, a sagitta $\lapproxeq 0.1$. However, there
 are complications (~see Appendix~).
\begin{figure}[htb!]
\begin{center}
\includegraphics[height=15cm,width=8cm,angle=-90]{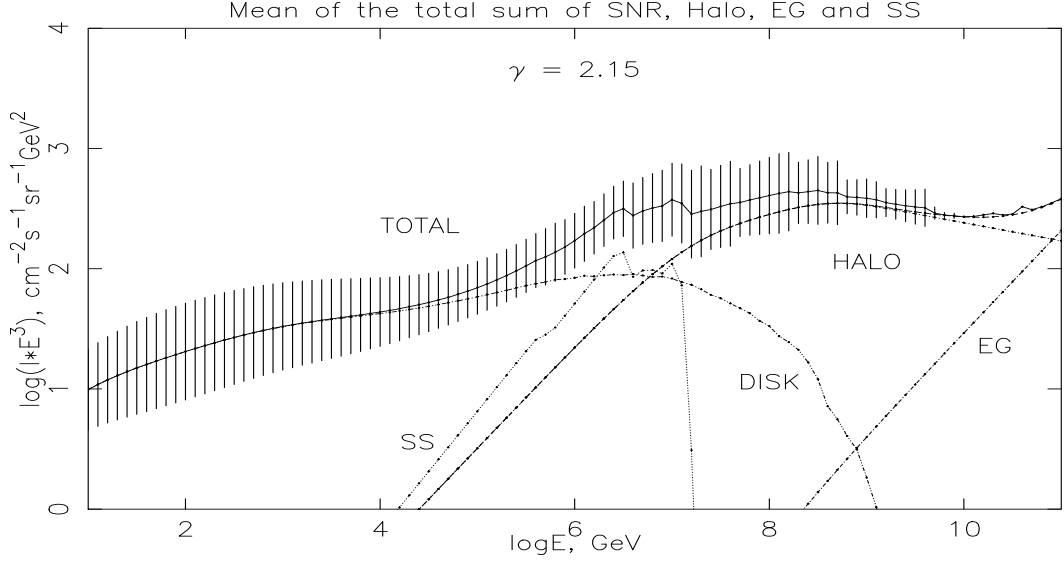}
\caption{\footnotesize The all particle CR spectrum as the sum of four
components, similar to that shown in Figure 7b, but for the emission spectrum in the
Halo, having slope index $\gamma = 2.15$. The irregularity of the total spectrum is
 substantially higher than for $\gamma = 2.7$.}
\end{center}
\label{fig:beyond8}
\end{figure}

\section{Conclusion}

Although the problem of the origin of particles with energy in
PeV-EeV region can hardly be said to be solved, there appear to be
two ways of injecting CR beyond the knee and up to a rigidity of
$logE = 8.5$ into the 'CR pool': very energetic SNR or young
pulsars with subsequent trapping in a Giant Halo (~size of $\sim
100$kpc~) with, or without subsequent acceleration. A 'long shot'
is the dominance of a small number of sources such as the pulsar
B0656+14 inside the Monogem Ring SNR which by chance give few
irregularities
 caused by the stochastic nature of multiple sources. However, as for pulsar B0656+14,
extended trapping of particles above PeV energies inside the
Monogem Ring shell is needed.

We should like to stress that, independently of their actual
origin, {\em CR particles above PeV energies most likely come
predominantly from the Galactic Halo with extended trapping}. It
is assumed that the Galactic Wind and its termination shock
provide conditions for such trapping. The small irregularity of
the observed CR energy spectrum, as well as the lack of large
anisotropies in arrival directions and the small
 radial gradient of CR, means that {\em the contribution of the Giant Halo to the total
 CR intensity is substantial at all energies up to about $10^{10} GeV$.} The leakage
from the Halo of our and other galaxies is the source of EG CR,
with some of these galaxies (~{\em eg} AGN~) being much more
powerful sources of CR than our Galaxy.

{\bf \large{Acknowledgments}}

The authors thank The Royal Society and the University of Durham for financial support.
A.R.Bell and H.J.V\"{o}lk are thanked for helpful comments.

\newpage
\begin{center}
{\bf \large{Appendix}}\\

{\bf \large{The relevance of variations of spectral shape}}
\end{center}
It is apparent that the fluctuations in spectral shape predicted
above the knee for at least some of the models considered are
dramatic and this provides the main case for postulating a
Galactic Halo with, or without, acceleration. At lower energies -
where direct measurements of individual nuclear charges have been
made - there are also significant fluctuations in shape predicted
in going from one 'run' (~a particular, randomly chosen set of SN
distances and ages~) to another. We examine this aspect in some
detail, starting with the results, and 'predictions', for 'heavy'
nuclei. An energy range $3\cdot 10^{10}$eV to $3\cdot 10^{12}$eV
has been chosen over which to study the spectral shapes insofar as
direct measurements have been made in this range.  The choice of
$\pm$ one decade in energy over which to search for variations of
spectral shape comes from two factors:

(i) Inspection of our predicted spectra (e.g. Figures 1 and 2
shows that `curvature' is often present on scales of the order of
a decade).

(ii) Experimental measurements are subject to both random and
systematic errors.  We consider that the random errors indicate
that wide bins of energy should be taken.  We consider that
systematic errors will usually cause slowly varying changes in
intensity which will only rarely contribute spurious curvature in
the spectrum.

As a first approach we determine the sagitta, $\delta$, of the difference between the
intensity at $3\cdot 10^{11}$eV and that found by linear (~in the logarithm~)
interpolation between the limiting energies (~$3\cdot 10^{10}$eV and
$3\cdot 10^{12}$eV~).

Two aspects are studied:

(i) The evidence for curvature in the spectrum, i.e. a finite value of $\delta$; the
basic shock acceleration model adopted by us \cite{Bere} shows increasing negative
curvature as the time of acceleration proceeds and observation of the correct curvature
 would give confidence in the model.

(ii) The likelihood, or otherwise, of each model providing an acceptable fit to the
data in the sense of not producing too high a value of $\delta$.

 We start with the experimental evidence for $Z \geq 2$; the results are shown in
Figure A1. The sources of the data are given in the caption to the
Figure. It is evident that there is a measure of consistency
between the different elements and that $\langle \delta \rangle$
is negative. Specifically, the mean is $\langle \delta
\rangle_{obs} = -0.11 \pm 0.03$.  It will be noted that there are
3 observations (out of 13) further than one sigma from the mean,
in comparison with an expected number of 4.

\begin{figure}[htb!]
\begin{center}
\includegraphics[height=12cm,width=8cm, angle=-90]{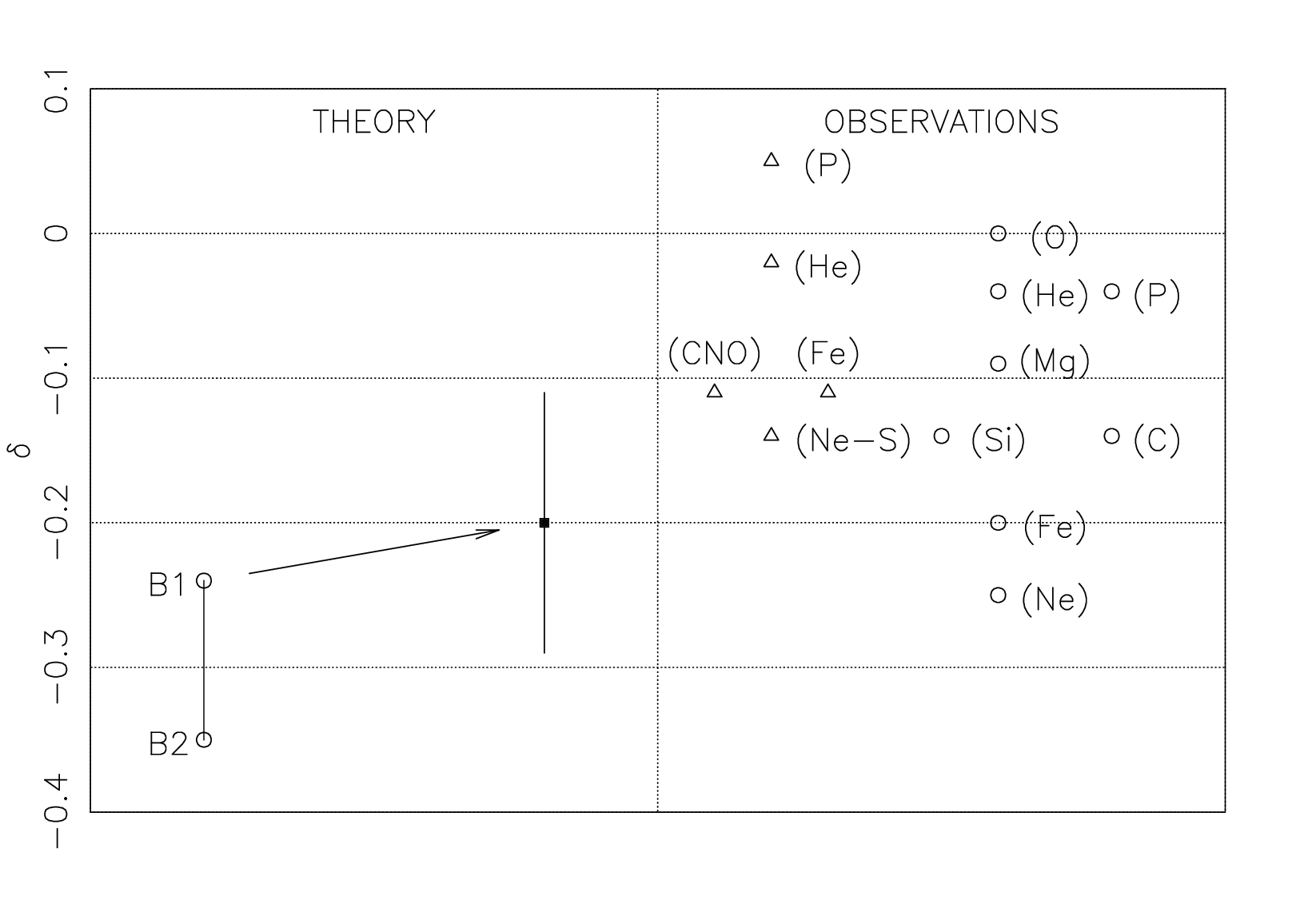}
\end{center}
{\footnotesize Figure A1: Sagittas (~$\delta$ - the spectral shape
factor~) for the range $3\cdot 10^{10} - 3\cdot 10^{12}$
eV/nucleon. Key: 'THEORY'. Shock acceleration model \cite{Bere}.
B1: 'efficiency', $\eta = 10^{-2}$; B2: $\eta = 10^{-4}$. The
value arrowed is after allowance for SNR of different properties
and environments. 'OBSERVATIONS'. $\bigtriangleup$ - \cite{Zats1},
$\bigcirc$ - \cite{Zats2}. The uncertainties on the points are,
typically $\pm 0.1$.} \label{fig:beyondA1}
\end{figure}

Turning to expectation, the shock acceleration model predicts a spectral shape at our
adopted time of SNR particle release (~$8\cdot 10^4$y~) which depends on the parameters
 of the SNR and ISM. Results for two values of the injection efficiency
(~$\eta = 10^{-4}$ and $\eta = 10^{-2}$~) are given in the Figure. Unlike the situation
 at higher energies, where there is a dramatic upturn (~in the $E^2 I(E)$ plot~), the
sagittas here are quite small.

Allowance for the frequency distribution of all the SNR and ISM properties is necessary
and an approximate estimate has been made; surprisingly, perhaps, the reduction of
$\delta$ is rather small. This is in contrast to the situation near the maximum energy
where the smoothing introduced is much more marked and the observed knee cannot be
reproduced with many different SNR. It is at the knee energy (~3 PeV~) where a 'Single
Source' is needed. At the energies concerned in this section
(~$3\cdot 10^{10} - 3\cdot 10^{12}$eV~) we are dealing with the 'background' spectrum
derived from many SNR.

We note that the observed sagitta, with mean $-0.11\pm 0.05$, is not inconsistent with
the expected shock model prediction: -0.1 to -0.3, although the range of predictions
is, admittedly, wide. What can be said, however, is that there is support for 
'heavy' nuclei, at least, being accelerated by the shock process insofar as curvature
seems to be a property of the mechanism.

Turning to comparison with our model predictions given in the present work, this is
not straightforward in the sense that a simple $E^{-2.15}$ (~or $E^{-2.7}$~) injection
spectrum was adopted, i.e. the mean $\delta$
is equal to zero. Thus, we need to make a systematic shift to the mean of the derived
sagittas for the constituent spectra. What is important is the spread in sagittas from
run to run. Figure A2 shows the frequency distribution for a series of situations, as
indicated in the caption.
\begin{figure}[htb!]
\begin{center}
\includegraphics[height=10cm,width=10cm]{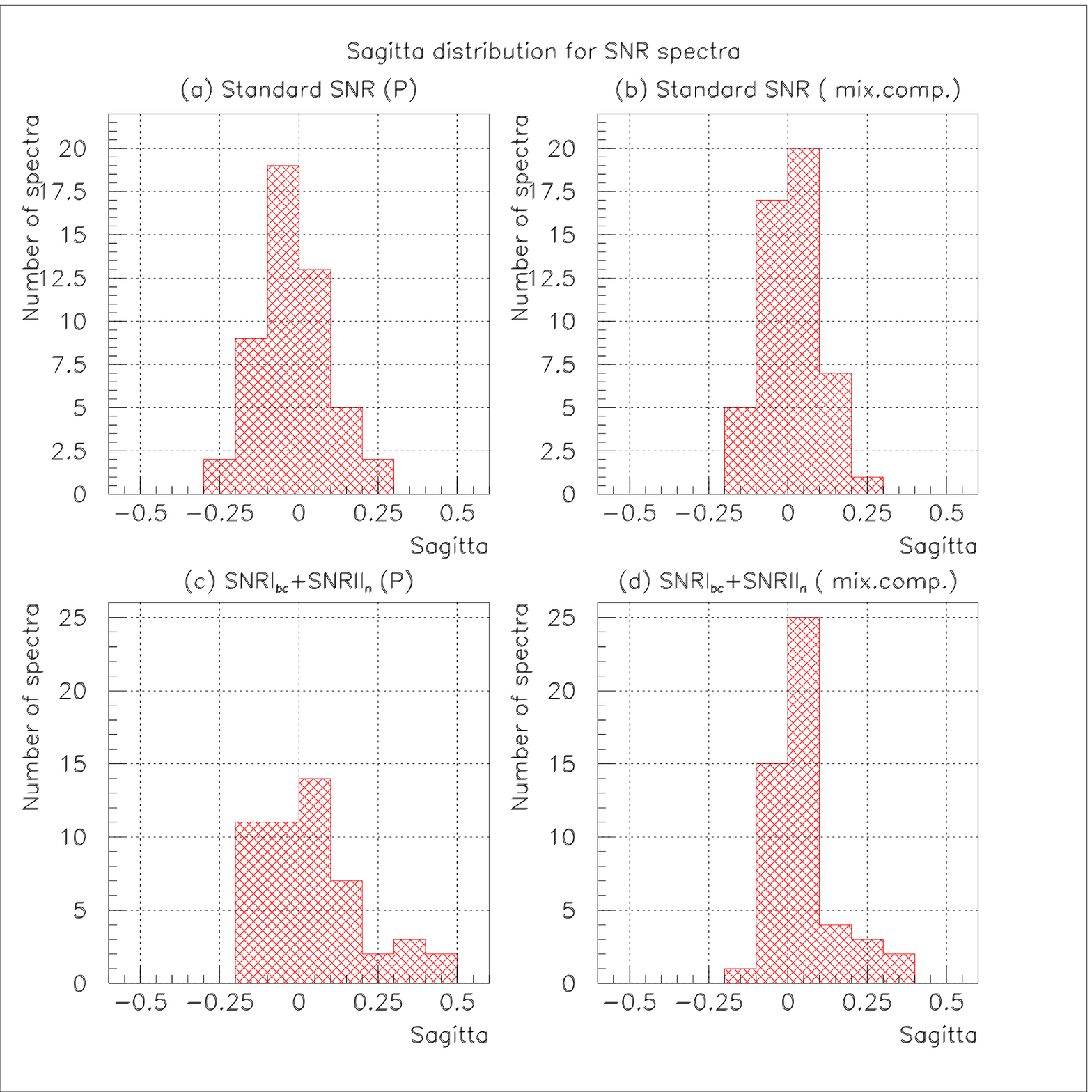}
\end{center}
{\footnotesize Figure A2: Distribution of sagittas of the energy
spectra for the range $3\cdot 10^{10} - 3\cdot 10^{12}$
eV/nucleon: (a) standard SNR, protons, $\langle \delta \rangle =
-0.018, rms = 0.112$; (b) standard SNR, mixed composition,
$\langle \delta \rangle =  0.014, rms = 0.091$; (c)
SNRI$_{bc}$+SNRII$_n$, protons; $\langle \delta \rangle =  0.04,
rms = 0.16$; (d) SNRI$_{bc}$+SNRII$_n$, mixed composition,
$\langle \delta \rangle = 0.048, rms = 0.103$.}
\label{fig:beyondA2}
\end{figure}
As remarked already, systematic displacement is necessary to allow
for a concave injection spectrum. The result is summarized in
Figure A1, where it will be noted that the situation with the
model in which there are 'super-SNR' and no Halo gives a spread
about twice that 'expected' :$\sigma = 0.16$ compared with our
observed $0.1\pm0.03$, a 2-sigma effect. With a Giant Halo,
however, and with a Galactic emission spectrum having $\gamma =
2.7$ (~Figure 7~) the situation is more agreeable (~$\sigma =
0.07$ compared with 0.1~). Before making a final conclusion with
respect to the sagitta method, some discussion for protons is
necessary. Here, for $E: 3\cdot 10^{10} - 3\cdot 10^{12}$eV, we
have $\langle \delta \rangle = 0.0 \pm 0.05$ (from 2 observations)
, viz., smaller than for the heavier nuclei. Insofar as the
measurements for protons extend to higher energies than for the
heavier particles, we can determine $\delta$ over a larger energy
range. The result is, for the range $10^{11} - 10^{14}$eV :
$\delta_P = +0.10 \pm 0.06$ Such a result is surprising, in that
for this enhanced energy range the value should be larger
(~negative~) than for the lower energy range if
 shock acceleration is valid. Two possibilities spring to mind:

$\ast$ the measurements are imprecise;

$\ast$ the protons are drawn preferentially from regions of the ISM where the
temperature is low, the acceleration efficiency is low and the spectrum more nearly
straight.

The second possibility seems the more likely.

The situation with the sagittas is therefore rather confused:
observationally, for nuclei with $Z \geq 2$ we have $\delta =
-0.11 \pm 0.03$ and for protons, $\delta = 0.0 \pm 0.05$, both for
$logE = 2.5 - 4.5$. Theoretically, for shock acceleration we
expect $\delta \sim -0.2 \pm 0.1$. Thus, it looks as though a
contribution of $\sim 0.1$ is required from stochastic processes.
Such a contribution could come from either $\gamma = 2.15$ or
$2.7$ although $\gamma = 2.7$ is preferred.

Moving to higher energies still, without a Giant Halo, the fluctuations increase rather
 quickly with energy
(~{\em eg} Figures 1 and 2~) and rapidly become too large. An
estimate of the observed sagitta for $logE: 8 - 11$ is $\delta =
-0.2 \pm 0.1$. Extrapolation of the results of Figure 2d,
corresponding to an inclusion of BSNR, leads to a $\sigma$-value
of $\simeq 1.6$, i.e. very much larger than observed. Indeed the
negative value of $\delta$ seen is usually associated with the
'ankle', above which extragalactic particles rapidly dominate.
This is particularly true above the knee when rare,
ultra-energetic, SNR or pulsars are invoked to endeavour to
explain the intensity. A Giant Halo appears to be necessary to
smooth the spectra. If we invoke acceleration at the termination
of the shock (~at $R\sim 100$kpc~) as the source of the 'Halo'
particles (~{\em eg} Figure 5~) then their flux - which will have
less variability - will also dilute the resulting spectral
fluctuations considerably.

The final magnitude of the fluctuations expected for the Halo
component is not easy to calculate, but an estimate can be made. A
number of factors are important, including variability in the
local Galactic Wind which reduces the intensity of the lower
energy particles (~$E\leq 1$PeV~) which are endeavouring to
re-enter the Galaxy, and, at all energies, the fluctuating,
wave-like nature of the shocks \cite{Voelk}. The last mentioned
have an amplitude in velocity of $\sim \pm$12\% at the boundary;
insofar as the energy carried, and thus transferred to CR, goes as
the square of the velocity, there is a 24\% temporal variability.
The original source of the energy, the Galactic Wind, is variable
not only in time but also in space, being greater above the spiral
arms than in the inter-arm regions. Galactic trapping will smooth
the ensuing fluctuations to a large extent and it seems unlikely
that there will be sagittas, measured over two orders of magnitude
for the range $logE$: 6 - 9, of more than about 10\%, i.e. $log
\delta = 0.04$. As Figure 4 indicates, this is allowable.

\begin{tabbing}
Table 1. \= {\bf Factors relevant to a Galactic Halo} \cite{Voelk} \\
{\bf Emission from the Galactic Disk \hspace{5cm}}\= {\bf Luminosity (~$erg \cdot s^{-1}$~)} \\
Cosmic Rays from the 'usual' SNR \> $3\cdot 10^{40}$ \\
(~$10^{50}$erg per SNR, rate $100^{-1}y^{-1}$~) \\
Starlight \> $4\cdot 10^{43}$ \\
(~$10^{10}$ stars, each of 1 $L_\odot$~) \\
Galactic Wind \> $3\cdot 10^{40}$ \\
(~$1 M_\odot y^{-1}$ with velocity $300 kms^{-1}$~) \\
{\bf Emission from the Galactic Disk \hspace{5cm}}\= {\bf Energy (~$erg s^{-1}$~)} \kill
{\bf Galactic Halo} \\
(~radius 100 kpc, energy independent lifetime $10^8$y (~or $10^{10}$y~) \\
{\bf 1. No Acceleration} \> {\bf Energy density} (~$eVcm^{-3}$~) \= \\
Energy density of CR escaping from the Disk \> $1.5\cdot 10^{-3}$ \hspace{5mm}  (0.15) \\
Energy density in the wind \> $1.4\cdot 10^{-2}$ \hspace{5mm} (1.4)\\
Energy density in the rotating wind \> $2.5\cdot 10^{-3}$ \hspace{5mm} (0.25)\\
{\bf 2. Acceleration at the Termination Shock} \\
Energy density of CR (~and gas and field, all approximately equal~), each: \\
Mean over 100 kpc \> $1.5\cdot 10^{-2}$ \hspace{5mm} (1.5)\\
Value at 100 kpc  \> $6\cdot 10^{-4}$ \hspace{8mm} ($6\cdot 10^{-2}$)\\
Energy density of CR above 1 PeV, pervading the Halo  \> $10^{-4}$ \hspace{13mm}  ($10^{-2}$)\\
(~Energy density needed to fit observations of the CR spectrum~) \> $10^{-4}$ \hspace{13mm}  ($10^{-2}$)\\
\end{tabbing}

\end{document}